\documentclass[letterpaper,aps,prd,twocolumn,floats,floatfix,amssymb,amsfonts,amsmath,superscriptaddress,showpacs]{revtex4}

\usepackage{graphicx}

\newcommand{\dd}[0]{\mathrm{d}}
\newcommand{\ee}[0]{\mathrm{e}}

\begin{document}

\title{AMS-02 data confronts acceleration of cosmic ray secondaries in
  nearby sources}
\author{Philipp Mertsch}
\affiliation{Kavli Institute for Particle Astrophysics \& Cosmology,
  2575 Sand Hill Road, M/S 29, Menlo Park, CA 94025, USA}
\author{Subir Sarkar}
\affiliation{Rudolf Peierls Centre for Theoretical Physics, University
  of Oxford, 1 Keble Road, Oxford OX1 3NP, UK} 
\affiliation{Niels Bohr Institute, Copenhagen University, Blegdamsvej
  17, 2100 Copenhagen \O, Denmark}
\begin{abstract}
We revisit the model proposed earlier to account for the observed
increase in the positron fraction in cosmic rays with increasing
energy, in the light of new data from the Alpha Magnetic Spectrometer
(AMS-02) experiment. The model accounts for the production and
acceleration of secondary electrons and positrons in nearby supernova
remnants which results in an additional, harder component that becomes
dominant at high energies. By fitting this to AMS-02 data we can
calculate the expected concomitant rise of the boron-to-carbon ratio,
as well as of the fraction of antiprotons. If these predictions are
confirmed by the forthcoming AMS-02 data it would conclusively rule
out all other proposed explanations, in particular dark matter
annihilations or decays.
\end{abstract}
\date{\today}
\maketitle

\section{Introduction}

Recently the AMS-02 collaboration have presented precision data on
cosmic ray (CR) protons, helium, electrons, positrons and the
boron-to-carbon ratio from the first two years of the space
mission~\cite{Aguilar:2013gta,AMS02:2013}. Some intriguing results
from earlier experiments have \emph{not} been corroborated, {\it e.g.}
there seems to be no `break' in the proton and helium spectra at $\sim
200$~GeV/n as had been claimed earlier~\cite{Adriani:2011cu}. However
the finding by PAMELA~\cite{Adriani:2008zr} of a rise in the positron
fraction with energy above $\sim 10$~GeV has been spectacularly
confirmed \cite{Aguilar:2013qda}.

This is of particular interest as the hardening of the positron
fraction had been widely interpreted as due to the annihilation
\cite{Bergstrom:2008gr} or decay \cite{Nardi:2008ix} of weak-scale
dark matter (DM). Such interpretations, while very exciting as
potential findings of new physics beyond the Standard Model, have
faced intrinsic challenges, {\it e.g.} the (velocity-averaged) DM
annihilation cross-section is required to be much larger than the
typical value which yields the observed DM abundance for a thermal
relic. Moreover the expected antiprotons are \emph{not} seen so the
annihilations or decays must be only into leptons which is rather
unnatural.  Subsequently, more direct constraints have been presented
on the associated energy
release~\cite{Ackermann:2012rg,Madhavacheril:2013cna}, severely
constraining DM interpretations. Astrophysical explanations
(see~\cite{Serpico:2011wg} for a review) have therefore gained more
currency with nearby pulsar wind nebulae being frequently implicated.

An interesting alternative suggestion is that a hard spectrum of
\emph{secondary} positrons can be produced by the standard sources of
Galactic cosmic rays (GCRs), i.e. supernova remnants
(SNRs)~\cite{Blasi:2009hv,Ahlers:2009ae} (see
also~\cite{Fujita:2009wk} for a related suggestion). This does not
require a new class of sources and has the added advantage that it is
easily falsifiable due to related signatures in other secondary
species, {\it e.g.} a rise is also predicted at higher energies in the
antiproton-to-proton ratio ($\mathrm{\bar{p}/p}$)~\cite{Blasi:2009bd},
and the boron-to-carbon (B/C)
ratio~\cite{Mertsch:2009ph,Tomassetti:2012ir}.

Until recently, such tests were hampered both by the lack of precision
in CR data and also the inconsistency between different data sets. In
this Letter we consider only the recently presented AMS-02 data which
have not only unprecedented statistics but also the smallest ever
systematic uncertainties. Besides fitting to the B/C and $e^\pm$ data
we present our model prediction for the $\mathrm{\bar{p}/p}$ ratio.
We improve on earlier studies by computing all observables
consistently, e.g. using the same nuclear cross-sections for the
source and propagation part of the calculation.  (See,
e.g.~\cite{Donato:2001ms} for a discussion of antiproton production
cross-sections.)

The remainder of this paper is organised as follows: In
Sec.~\ref{sec:accsec} we compute the contribution from the production
and acceleration of secondaries in the source, i.e. supernova
remnants. For the transport of all CR species in the ISM we use the
\texttt{GALPROP} code and we explain our approach in
Sec.~\ref{sec:transport}. In Sec.~\ref{sec:results}, we present our
results for the positron fraction, the lepton fluxes, proton and
helium fluxes as well as B/C. In addition, for the
antiproton-to-proton ratio where data from AMS-02 is expected soon, we
compare to currently available PAMELA data. We conclude in
Sec.~\ref{sec:conclusion}.

\section{Production and acceleration of secondaries in the source}
\label{sec:accsec}

It is generally believed that collisionless shock waves in supernova
remnants (SNRs) are the dominant agent for acceleration of GCRs
\cite{Blasi:2013hoa}. After the shock cannot contain particles anymore
they diffuse through the interstellar medium (ISM), producing
secondary particles by spallation on the interstellar gas. The
production of secondary particles \emph{inside} SNRs has largely been
ignored (see, however,~\cite{Berezhko:2003pf}) since the total
grammage of ambient matter that primary CRs traverse therein is much
smaller than the grammage they traverse in the ISM. However, it was
realised~\cite{Blasi:2009hv,Ahlers:2009ae} that charged secondaries
like positrons, anti-protons or boron nuclei partake in shock
acceleration in much the same way as their parent primaries. However,
whereas primaries are injected from the background thermal plasma only
at the shock, secondary particles are produced up to $\mathcal{O}(1)$
diffusion scale length away. This leads to a different spatial
distribution for their injection and is reflected in a secondary
spectrum \emph{harder} than the primary one due to the energy
dependent diffusion coefficient. Therefore, although subdominant in
total number, secondaries produced in the SNR can have observable
consequences at high enough energies.

Here, we consider the acceleration of primary and secondary CRs in the
test-particle approximation of diffusive shock acceleration. In its
own rest-frame, the shock is at $x=0$ and the Rankine-Hugoniot
conditions determine the compression factor $r$ which fixes the ratio
of gas densities and velocities $n_{+}/n_{-} = r = u_{-}/u_{+}$ on
either side of the shock. The evolution of the gyro-phase and
pitch-angle averaged phase space density $f_i \equiv f_i(x,p)$ of
species $i$ is governed by the transport equation:
\begin{equation}
\frac{\partial f_i}{\partial t} = -u\frac{\partial f_i}{\partial x} +
\frac{\partial}{\partial x}D_i \frac{\partial f_i}{\partial x} -
\frac{p}{3} \frac{\dd u}{\dd x} \frac{\partial f_i}{\partial p} -
\Gamma_i f_i + q_i \,,
\label{eqn:transport}
\end{equation}
where from left to right, the terms on the right hand side describe
convection, spatial diffusion, adiabatic losses/gains, inelastic
losses and injection by spallation of heavier species, $q_i =
\sum_{j>i} c \beta_j n_{\text{gas}} \sigma_{j \rightarrow i} f_j$.

We solve for the steady state solutions, \mbox{$f_i^{\pm} \equiv f_i(x
  \gtrless 0)$}, separately in the upstream ($x < 0$) and downstream
($x > 0$) regions where $\dd u / \dd x \equiv 0$, and impose the
boundary conditions, $f_i^{+} < \infty$ and $\partial f_i^{-} /
\partial x \rightarrow 0$ for $x \rightarrow \infty$ as well as
$f_i^{-} \rightarrow Y_i \delta (p-p_0)$ for $x \rightarrow - \infty$,
where $Y_i$ is the injected abundance of species $i$ and $p_0$ the
injection momentum. Assuming that $\Gamma_i D_i / u_{\pm}^2 \ll 1$ and
$x \Gamma_i / u_{\pm} < x_{\text{max}} \Gamma_i / u_{\pm} \ll 1$
(which amounts to requiring efficient acceleration of
nuclei~\cite{Mertsch:2009ph}), we find for the downstream solution:
\begin{equation}
f_i^{+}(x,p) = f_i^0(p) + r (q_i^0(p) - \Gamma_i^- f_i^0(p)) \frac{x}{u_{+}}\, ,
\end{equation}
where $q_i^0(p) \equiv q_i^{-}(x=0,p)$ is the upstream injection term
at the shock, and
\begin{align}
f_i^0(p) = & \int_0^p \frac{\dd p'}{p'} \left(\frac{p'}{p}\right)^{\gamma} 
\ee^{-\gamma(1+r^2)\left(D_i(p) - D_i(p')\right)\Gamma_i^{-}(p)/u_{-}^2} \nonumber \\ 
& \times \left[\gamma (1+r^2)\frac{D_i(p')}{u_{-}^2} q_i^0(p') 
 + \gamma Y_i \delta(p'-p_0)\right]
\label{eqn:fi0}
\end{align}
is the phase space density at the shock. Without spallation and
inelastic losses, {\it i.e.} for $q_i = 0, \Gamma_i = 0$, the
well-known test-particle solution of diffusive shock acceleration,
$f_i \propto p^{- \gamma}$, is recovered, with the spectral index
$\gamma = 3r/(r-1)$, i.e. $\gamma = 4$ for a strong shock ($r =
4$). For non-zero spallation and assuming that the diffusion
coefficient is proportional to momentum $D_i(p) \propto p$ ({\it i.e.}
Bohm diffusion), $f_i^0(p)$ will be harder than the source spectrum $
q_i^0(p)$ by one power in momentum. This results in an \emph{increase}
of the fraction of positrons with energy, and this will also be the
case for other secondary species like boron or antiprotons.

We make the simplifying assumption that after a time
$\tau_{\text{SNR}}$, the effective lifetime of the SNR, all
down-stream particles are released in a time much shorter than the
time needed for the particles to reach the observer at Earth. The
integrated down-stream spectrum is:
\begin{align}
\frac{\dd N_i}{\dd p} &= 4\pi\int_0^{\tau_{\text{SNR}} u_{+}} \dd x \, 
x^2 4\pi p^2 f_i(x,p) \nonumber \\ 
&= 4\pi p^2 V \left[f_i^0 + \frac{3}{4}r \,\tau_{\text{SNR}} 
\left(q_i^0 - \Gamma_i^- f_i^0 \right)\right]\, ,
\label{eqn:TotalDownstreamSpectrum}
\end{align}
where $V = \frac{4\pi}{3}(\tau_{\text{SNR}}u_{+})^3$ is the downstream
volume. We note that the term $-\frac{3}{4}r\tau_{\text{SNR}}
\Gamma_i^{-}f_i^0(p)$ in eq.(\ref{eqn:TotalDownstreamSpectrum}) as
well as the exponential in eq.(\ref{eqn:fi0}) will lead to a
suppression of the secondary contribution at very high energies.

\section{Transport of Galactic Cosmic Rays}
\label{sec:transport}

For the transport of CRs in the ISM we employ the \texttt{GALPROP}
code~\cite{Moskalenko:1997gh} which numerically solves a transport
equation similar to eq.(\ref{eqn:transport}) but with the total
downstream spectrum from eq.(\ref{eqn:TotalDownstreamSpectrum}) as the
source term and with transport parameters (diffusion coefficient, gas
densities, energy losses) appropriate for the ISM. The spallation on
interstellar gas of primary CRs, which are already softer than the
source spectrum due to escape losses, leads to further injection of
secondaries. These secondaries themselves suffer escape losses and are
therefore further softened. At low energies, where the secondaries
produced and accelerated in the SNRs are subdominant,
secondary-to-primary ratios such as the positron fraction, B/C and
$\mathrm{\bar{p}/p}$, are therefore expected to fall with energy, as
is in fact observed. However at higher energies the harder secondaries
come to dominate and the secondary-to-primary ratios should start to
\emph{rise} with energy.

Given that SNRs occur at random in the Galaxy, the flux from a
distribution of burst- and point-like sources will in general differ
from the flux assuming a smooth source density. This is particularly
important for high energy electrons and positrons which have limited
propagation lengths due to synchrotron and inverse Compton losses. We
have therefore performed the propagation of light nuclei and leptons
in the 3-dimensional, stochastic SNR mode of \texttt{GALPROP} and
recorded the fluxes for a statistical ensemble of 25 different
realisations of a pulsar-like~\cite{Lorimer:2003qc} source
distribution. For lepton fluxes, the envelope of the fluxes is shown
by the shaded bands in the following figures while for nuclei they are
sufficiently narrow and are therefore suppressed --- we show this for
illustration for the proton flux alone.

\section{Results}
\label{sec:results}

There are several free parameters in our model that determine the
source spectra, namely $r$, $u_{\pm}$, $\tau_{\text{SNR}}$,
$n_{\text{gas}}$ as well as the diffusion coefficient $D = \beta c
r_\text{L}(p)/3 \simeq 3\times 10^{22}\,K_\text{B}\,(p
c/\text{GeV})\,Z^{-1} B_{\mu \text{G}}^{-1}\, \text{cm}^2
\,\text{s}^{-1}$, where $r_\text{L}(p)$ is the Larmor radius. Here,
$K_\text{B} \sim B^2/\delta B^2$ parametrises deviations from the Bohm
value, arising \emph{e.g.} because at late stages of the SNR evolution
field amplification is less efficient. (Moreover, the adopted
test-particle limit is then a good approximation.)  However, of
these parameters, only the combination $K_\text{B} /(u_{-}^2 B)$
enters into the secondary terms, so we fix $B_{\mu\text{G}} = 1$ and
$u_{-} = 5 \times 10^7\,\text{cm}\,\text{s}^{-1}$, values typical of
old SNRs, and vary only $K_\text{B}$. Similarly we fix $n_{\text{gas}}
= 2\,\text{cm}^{-3}$ and test different values of $\tau_{\text{SNR}}$.

In choosing the parameters that describe the propagation, we cannot
rely on studies which do \emph{not} consider the contribution from
secondaries as this can be important even at the lowest energies for
the B/C or $\mathrm{\bar{p}/p}$ ratio.  We therefore fit the relevant
parameters in the following order. First, varying the source
parameters $\tau_\text{SNR}$ and $D$ as well as the propagation
parameters $\kappa$ (the ISM diffusion coefficient at a reference
rigidity $4 \, \text{GV}$), $\delta$ (its spectral index) and $\dd v /
\dd z$ (the gradient of the galactic wind), we attempt to
simultaneously reproduce B/C and the positron flux. The proton
spectral index and normalisation are then fixed by fitting to the
AMS-02 proton flux. Nuclear abundances relative to protons are adopted
from earlier studies~\cite{Ptuskin:2005ax}, however fitting helium
data requires a spectral index $\gamma_{\text{He}}$ harder by $\sim
0.1$ compared to that of other nuclei, $\gamma$ which is a known
issue~\cite{Adriani:2011cu}. We fix the electron spectral index both
above and below a fixed break energy of $7 \, \text{GeV}$ as is
required by radio observations~\cite{Strong:2011wd}, and the
normalisation by fitting to the AMS-02 electron flux. The positron
fraction and $\mathrm{\bar{p}/p}$ are then \emph{predictions} of the
model. We adopt the force-field approximation of Solar
modulation~\cite{Gleeson:1968}, allowing for different modulation
potentials for the various species within the commonly adopted range
$0.2-0.8 \,\text{GV}$. The half-height of the diffusion volume,
$z_{\text{max}}$, is always fixed to $3 \,\text{kpc}$

We can thus fix most of the model parameters, however due to the
limited energy range of available data there remains some freedom
concerning the maximum energy $E_\text{max}$ (or equivalently rigidity
$R_\text{max}$). In DSA models where the maximum energy is age-limited
it would be determined by the diffusion coefficient and shock
velocity, but in models where it is escape-limited it is a complicated
function of the age of the source. Since we expect the biggest
contribution from mature supernova remnants where the diffusion
coefficient is relatively large (as magnetic field amplification is no
longer efficient) and the shock speed is relatively low (\emph{c.f.}
eq.~\ref{eqn:fi0}), we allow $E_\text{max}$ to range between 1 and
tens of TeV.

\begin{figure}[!t]
\includegraphics[scale=.93]{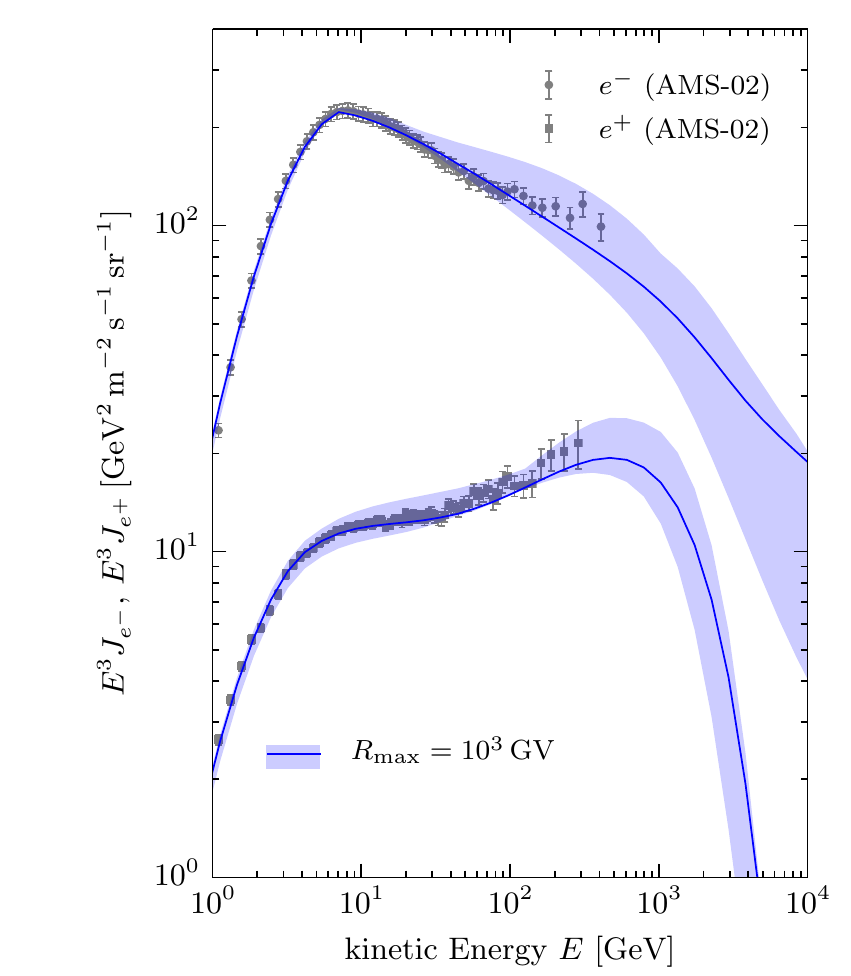}
\caption{Electron and positron fluxes measured by AMS-02
  (circles and squares, respectively) and for the acceleration of
  secondaries model with maximum rigidity of $1 \, \text{TV}$. The
  shaded band reflects the uncertainty of the spatial and temporal
  distribution of the SNRs.}
\label{fig:leptons1}
\end{figure}

\begin{figure}[t!]
\includegraphics[scale=.93]{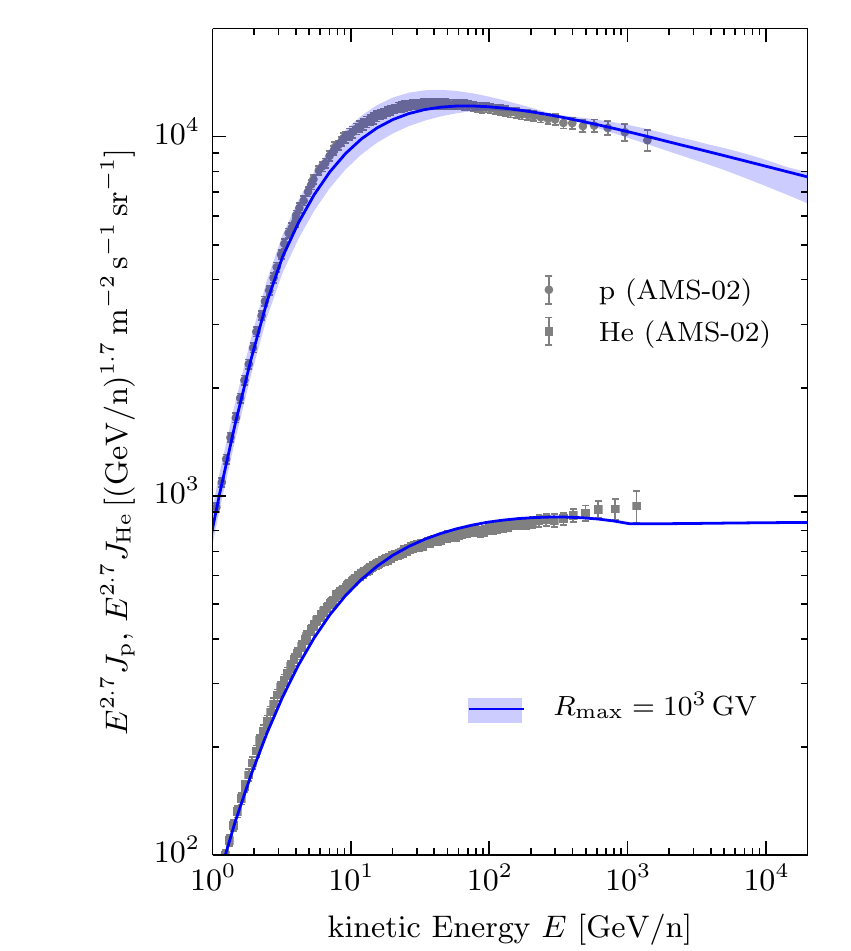}
\caption{Proton and helium fluxes measured by AMS-02 (circles
  and squares, respectively) and for the acceleration of secondaries
  model with maximum rigidity of $1 \, \text{TV}$. The shaded band
  reflects the uncertainty of the spatial and temporal distribution of
  the SNRs.}
\label{fig:nuclei1}
\end{figure}

\begin{figure}[!t]
\includegraphics[scale=.93]{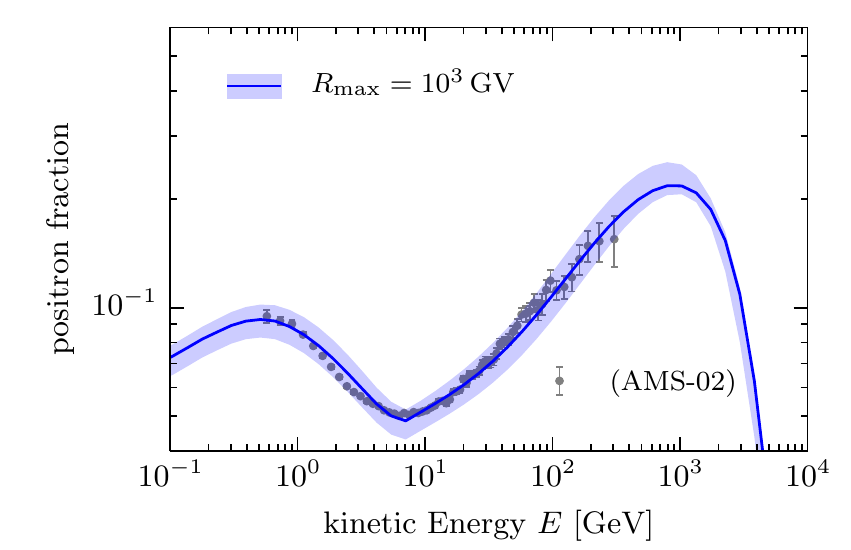}
\caption{The positron fraction, $e^+/(e^+ + e^-)$ measured by AMS-02
  (circles) and for the acceleration of secondaries model with maximum
  rigidity of $1 \, \text{TV}$. The shaded band reflects the
  uncertainty of the spatial and temporal distribution of the SNRs.}
\label{fig:PF1}
\end{figure}

We adopt a benchmark model with $R_\text{max} = 1 \, \text{TV}$. The
other model parameters adopted are shown in Tbl.~\ref{tbl:parameters}
(which also lists two other models we considered). In
Figs.~\ref{fig:leptons1}--\ref{fig:PF1} our results are compared with
AMS-02 data~\cite{AMS02:2013}.

\begin{table}[t]
\begin{tabular*}{\columnwidth}{@{\extracolsep{\fill} } l c c c}
\hline \hline
$R_\text{max}$ & 1 TV 		& 3 TV 		& 10 TV \\
$K_\text{B}$ 	& 16 			& 5 			& 8 \\
$\tau_\text{SNR} [10^4 \, \text{yr}]$ & 5 	& 4 	& 4 \\
$\gamma$ 	& 4.15 		& 4.05  		& 4.10 \\
$\gamma_{\text{He}}$ & 4.05 	& 3.95 		& 4.00 \\
$\gamma_{e,1}$ & 3.6 		& 3.6 		& 3.6 \\
$\gamma_{e,2}$ & 4.55 		& 4.50 		& 4.50 \\
\hline
$\delta$ 		& 0.65 		& 0.75 		& 0.70 \\
$\kappa$ [$10^{28}\,\text{cm}^2 \text{s}^{-1}$] & 2.80 & 2.00 & 2.10 \\
$(\dd v / \dd z)$ [$\text{km} \text{s}^{-1} \text{kpc}^{-1}$] & 5 & 15 & 15 \\
\hline \hline
\end{tabular*}
\caption{Parameter values of the models adopted in our analysis,
  both for the source ($K_\text{B}$, $\gamma$, $\gamma_{\text{He}}$,
  $\gamma_{e,1}$ and $\gamma_{e,2}$) and for the galactic propagation
  ($\delta$, $\kappa$ and $(\dd v/\dd z)$).}
\label{tbl:parameters}
\end{table}

\begin{figure}[!thb]
\includegraphics[scale=.93]{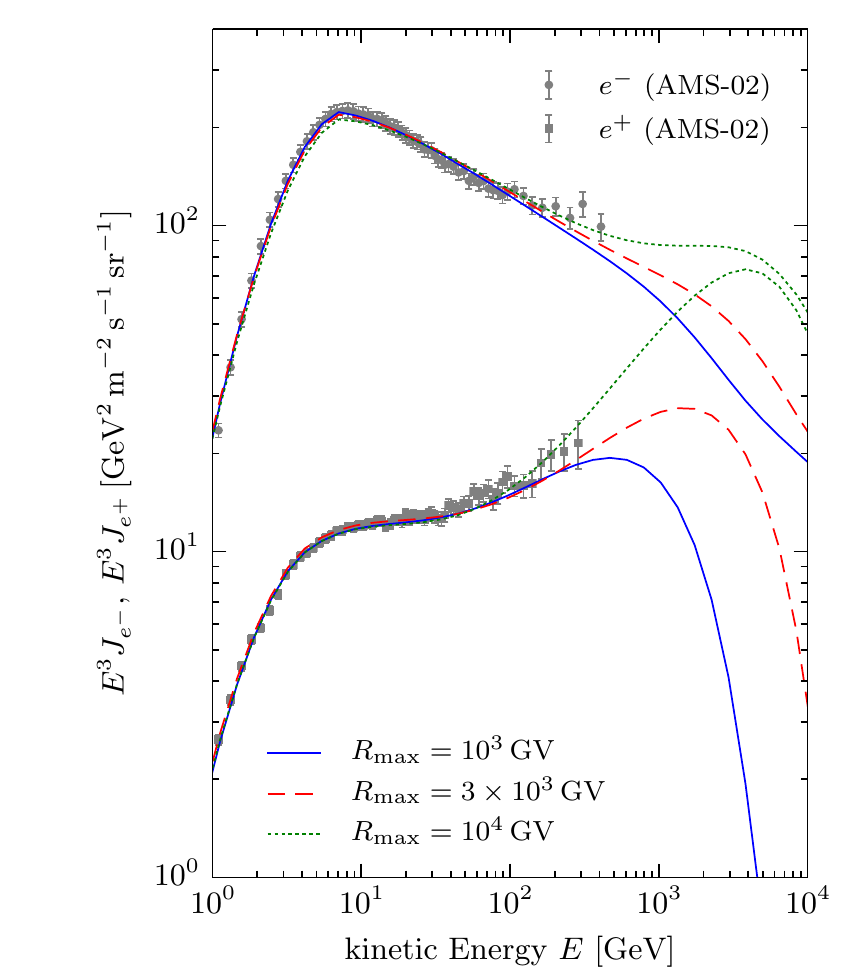}
\caption{Electron (circles) and positron (squares) fluxes
  measured by AMS-02, and predicted by the acceleration of secondaries
  model with maximum rigidities of $1$, $3$ and $10 \, \text{TV}$.}
\label{fig:leptons3}
\end{figure}

\begin{figure}[t!]
\includegraphics[scale=.93]{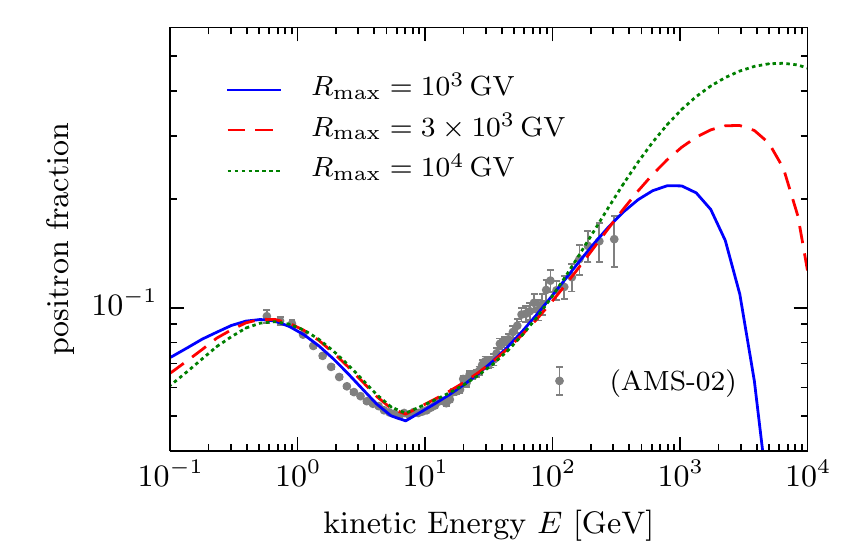}
\caption{The positron fraction, $e^+/(e^+ + e^-)$ measured by AMS-02
  (circles), and predicted by the acceleration of secondaries model
  with maximum rigidities of $1$, $3$ and $10 \, \text{TV}$.}
\label{fig:PF3}
\end{figure}

\begin{figure}[t!]
\includegraphics[scale=.93]{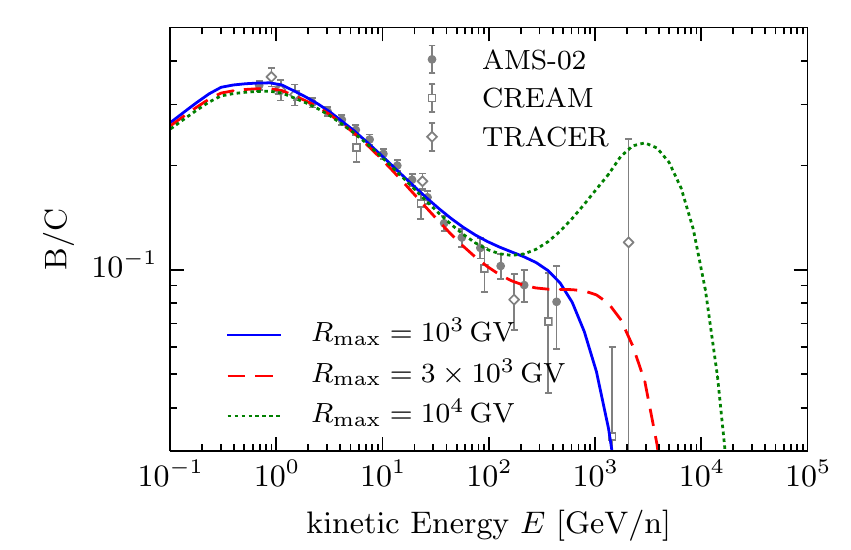}
\caption{The boron-to-carbon ratio measured by AMS-02 (circles), CREAM
  (open squares), TRACER (open diamonds), and predicted by the
  acceleration of secondaries model with maximum rigidities of $1$,
  $3$ and $10 \, \text{TV}$.}
\label{fig:B2C3}
\end{figure}

\begin{figure}[t!]
\includegraphics[scale=.93]{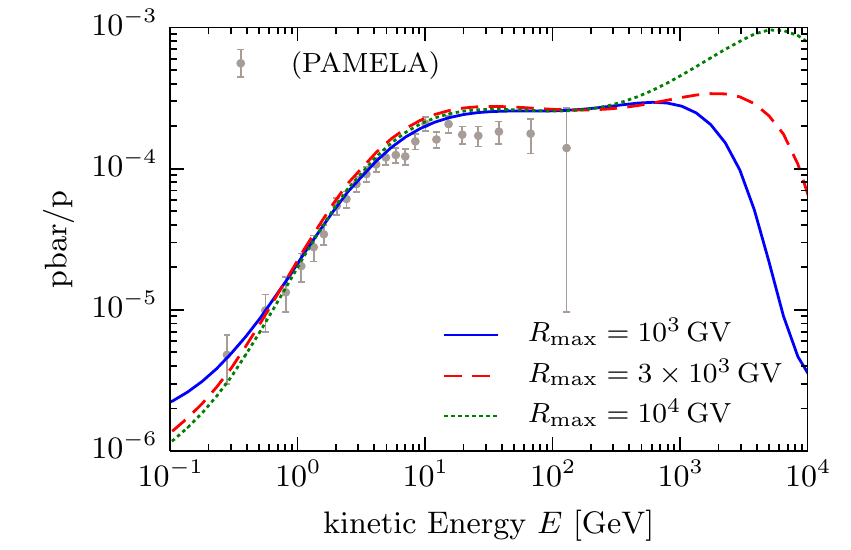}
\caption{The antiproton-to-proton ratio measured by PAMELA (circles),
  and predicted by the acceleration of secondaries model with maximum
  rigidities of $1$, $3$ and $10 \, \text{TV}$.}
\label{fig:pbar2p3}
\end{figure}

A point of difference with earlier
studies~\cite{Blasi:2009hv,Blasi:2009bd,Mertsch:2009ph,Ahlers:2009ae},
is that our model parameters are chosen to reproduce the shallower
rise of the AMS-02 positron fraction (compared to PAMELA or Fermi-LAT
data) at high energies which also allows a fit to the tempered rise in
the positron flux shown in Fig.~\ref{fig:leptons1}. We emphasise that
reproducing this as well as the electron flux shown in the same figure
is directly constrained by the fit to the proton and helium fluxes in
Fig.~\ref{fig:nuclei1}. As seen in Fig.~\ref{fig:PF1}, both models
provide good fits to the positron fraction measured by AMS-02.

To illustrate the spectral dependence on the maximum rigidity, we have
varied the latter in the range $1 \mathellipsis 10 \, \text{TV}$.
Figs.~\ref{fig:leptons3} and~\ref{fig:PF3} show our results for
$R_\text{max} = 3$ and $10 \, \text{TV}$ together with those shown
earlier for $R_\text{max} = 1 \, \text{TV}$. In Fig.~\ref{fig:B2C3},
we compare the AMS-02 measurements with our prediction for B/C; we
also show two recent balloon-borne measurements,
viz. CREAM~\cite{Ahn:2008my} and TRACER~\cite{Obermeier:2011wm}. This
displays the same behaviour as the positron fraction --- a fall at low
energies where the (softer) boron flux produced by CR primaries in the
ISM dominates, and a hardening at higher energies where the (harder)
flux of borons produced and accelerated inside SNRs dominates. We also
compare in Fig.~\ref{fig:pbar2p3} our antiproton-to-proton ratio to
PAMELA data~\cite{Adriani:2010rc}.

The other parameters for $R_\text{max} = 3$ and $10 \, \text{TV}$ are
shown in the third and fourth column of
Tbl.~\ref{tbl:parameters}. Note that while $R_\text{max}$ and
$K_\text{B}$ may be (anti-)correlated, this depends in detail on how
the cosmic rays escape from the SNRs --- in the absence of a firm
understanding we have treated these as independent parameters (and let
the fit to the data determine the value of $K_\text{B}$). Also we have
\emph{not} done a comprehensive scan of all the parameters, hence the
curves do not always vary monotonically.

Given their limited energy range and uncertainties, the presently
available electron and positron data (see Fig.~\ref{fig:leptons3})
cannot pin down $R_\text{max}$. However Fig.~\ref{fig:PF3} illustrates
that higher energy measurements of the positron fraction with better
statistics can distinguish between maximum rigidities of $1 \,
\text{TV}$ and tens of TV. This is an important point as
$R_\text{max}$ --- if it is the same for all secondary species as is
assumed here --- leads to qualitatively different behaviours for B/C:
While for $R_\text{max} = 1 \, \text{TV}$, B/C shows only a slight
hardening just below the cut-off, it flattens out for $R_\text{max} =
3 \, \text{TV}$, and even shows a characteristic rise for
$R_\text{max} = 10 \, \text{TV}$. The minimum in the latter case is
close to the highest energy bin for which AMS-02 have presented
data. Note that this minimum is at a \emph{different} energy for B/C
and for the positron fraction. This is due to the different kinematics
(positrons are on average produced at $\sim 1/20$ of the parent
primary energy, whereas in spallation the energy per nucleon is
roughly conserved), and also due to the spectral softening in the
primary electron spectrum.

However as seen in Fig.~\ref{fig:pbar2p3}, the $\mathrm{\bar{p}/p}$
fraction shows a flattening between tens and hundreds of GeV, unlike
the positron fraction or B/C. At these energies, the antiproton flux
is dominated by the secondary contribution which has the same spectrum
as the primary species (\textit{cf.} the term $4 \pi p^2 V (3/4 r
\tau_\text{SNR} q_i^0)$ in eq.~(\ref{eqn:TotalDownstreamSpectrum}); the
effect of the upper rigidity cutoff becomes apparent only at higher
energies where the term $4 \pi p^2 V f_i^0$ starts to dominate.

\section{Conclusion}
\label{sec:conclusion}

We have presented results for the (absolute) electron and positron,
proton and helium fluxes, as well as B/C and $\mathrm{\bar{p}/p}$, in
the framework of the acceleration of secondaries by SNR shock waves
model. The only free parameter that cannot be fixed by fitting to
available data from AMS-02 is the maximum rigidity to which cosmic
rays are accelerated by the SNR shock wave. Depending on whether it is
high ({\it e.g.} $10\,\text{TV}$) or low ({\it e.g.} $1\,\text{TV}$),
the positron fraction will keep increasing beyond a TeV, or cut off
shortly above the highest energy bin for which results have been shown
by AMS-02. This behaviour should be reflected by a cut-off or a rise
after a shallow minimum in B/C. For $\mathrm{\bar{p}/p}$, we have
found a plateau between tens and hundreds of GeV.

Our results differ significantly from Ref.\cite{Cholis:2013lwa} since
these authors fixed $\delta=0.43$ for the energy-dependence of the ISM
diffusion co-efficient, whereas we have considered larger values in
the range $\delta=0.65-0.75$ as is expected in diffusion-convection
models of CR transport \cite{Maurin:2010zp}. This is essentially why
we are able to consistently fit \emph{both} the positron fraction and
the B/C ratio. We await the release of AMS-02 data on the
$\mathrm{\bar{p}/p}$ and B/C ratio, which will definitively test all
models proposed to account for the rising positron fraction.

\section*{Acknowledgements}

We are grateful to Stefan Schael for helpful discussions. PM is
supported by DoE contract DE-AC02-76SF00515 and a KIPAC Kavli
Fellowship. SS acknowledges a DNRF Niels Bohr Professorship.


\begin{thebibliography}{99}

\bibitem{Aguilar:2013gta} 
  M.~Aguilar [AMS Collaboration],
  CERN Cour.\  {\bf 53}, no. 8, 23 (2013).

  
\bibitem{AMS02:2013}
  [AMS Collaboration], 
  Proc. 33rd Intern. Cosmic Ray Conf. (2013).

\bibitem{Adriani:2011cu} 
  O.~Adriani {\it et al.}  [PAMELA Collaboration],
  Science {\bf 332}, 69 (2011)
  [arXiv:1103.4055 [astro-ph.HE]].

\bibitem{Adriani:2008zr} 
  O.~Adriani {\it et al.}  [PAMELA Collaboration],
  Nature {\bf 458}, 607 (2009)
  [arXiv:0810.4995 [astro-ph]].
  
\bibitem{Aguilar:2013qda} 
  M.~Aguilar {\it et al.}  [AMS Collaboration],
  Phys.\ Rev.\ Lett.\  {\bf 110}, 141102 (2013).
  
\bibitem{Bergstrom:2008gr} 
  L.~Bergstrom, T.~Bringmann and J.~Edsjo,
  Phys.\ Rev.\ D {\bf 78}, 103520 (2008)
  [arXiv:0808.3725 [astro-ph]].

\bibitem{Nardi:2008ix} 
  E.~Nardi, F.~Sannino and A.~Strumia,
  JCAP {\bf 0901}, 043 (2009)
  [arXiv:0811.4153 [hep-ph]].

\bibitem{Ackermann:2012rg} 
  M.~Ackermann {\it et al.}  [LAT Collaboration],
  Astrophys.\ J.\  {\bf 761}, 91 (2012)
  [arXiv:1205.6474 [astro-ph.CO]].

\bibitem{Madhavacheril:2013cna} 
  M.~S.~Madhavacheril, N.~Sehgal and T.~R.~Slatyer,
  Phys.\ Rev.\ D {\bf 89}, 103508 (2014)
  [arXiv:1310.3815 [astro-ph.CO]].

\bibitem{Serpico:2011wg} 
  P.~D.~Serpico,
  Astropart.\ Phys.\  {\bf 39-40}, 2 (2012)
  [arXiv:1108.4827 [astro-ph.HE]].

\bibitem{Blasi:2009hv} 
  P.~Blasi,
  Phys.\ Rev.\ Lett.\  {\bf 103}, 051104 (2009)
  [arXiv:0903.2794 [astro-ph.HE]].

\bibitem{Ahlers:2009ae} 
  M.~Ahlers, P.~Mertsch and S.~Sarkar,
  Phys.\ Rev.\ D {\bf 80}, 123017 (2009)
  [arXiv:0909.4060 [astro-ph.HE]].

\bibitem{Fujita:2009wk} Y.~Fujita, K.~Kohri, R.~Yamazaki, K.~Ioka,
  K.~Kohri, R.~Yamazaki and K.~Ioka,
  Phys.\ Rev.\ D {\bf 80}, 063003 (2009)
  [arXiv:0903.5298 [astro-ph.HE]].

\bibitem{Blasi:2009bd} 
  P.~Blasi and P.~D.~Serpico,
  Phys.\ Rev.\ Lett.\  {\bf 103}, 081103 (2009)
  [arXiv:0904.0871 [astro-ph.HE]].

\bibitem{Mertsch:2009ph} 
  P.~Mertsch and S.~Sarkar,
  Phys.\ Rev.\ Lett.\  {\bf 103}, 081104 (2009)
  [arXiv:0905.3152 [astro-ph.HE]].

\bibitem{Tomassetti:2012ir} 
  N.~Tomassetti {\it et al.}  [ Collaboration],
  Astron.\ Astrophys.\  {\bf 544}, A16 (2012)
  [arXiv:1203.6094 [astro-ph.HE]].

\bibitem{Donato:2001ms} F.~Donato, D.~Maurin, P.~Salati, A.~Barrau,
  G.~Boudoul and R.~Taillet,
  Astrophys.\ J.\  {\bf 563}, 172 (2001)
  [astro-ph/0103150].

\bibitem{Blasi:2013hoa} 
  P.~Blasi,
  Astron.\ Astrophys.\ Rev.\  {\bf 21}, 70 (2013).

\bibitem{Berezhko:2003pf} E.~G.~Berezhko, L.~T.~Ksenofontov,
  V.~S.~Ptuskin, V.~N.~Zirakashvili and H.~J.~Voelk,
  Astron.\ Astrophys.\  {\bf 410}, 189 (2003)
  [astro-ph/0308199].

\bibitem{Moskalenko:1997gh} 
  I.~V.~Moskalenko and A.~W.~Strong,
  Astrophys.\ J.\  {\bf 493}, 694 (1998)
  [astro-ph/9710124].
 
\bibitem{Gleeson:1968}
  L.~Gleeson, W.~Axford,
  Astrophys.\ J.\  {\bf 154}, 1011 (1968).

\bibitem{Lorimer:2003qc} 
  D.~R.~Lorimer,
  astro-ph/0308501.

\bibitem{Ptuskin:2005ax} V.~S.~Ptuskin, I.~V.~Moskalenko, F.~C.~Jones,
  A.~W.~Strong and V.~N.~Zirakashvili,
  Astrophys.\ J.\  {\bf 642}, 902 (2006)
  [astro-ph/0510335].

\bibitem{Strong:2011wd} 
  A.~W.~Strong, E.~Orlando and T.~R.~Jaffe,
  Astron.\ Astrophys.\  {\bf 534}, A54 (2011)
  [arXiv:1108.4822 [astro-ph.HE]].

\bibitem{Ahn:2008my} H.~S.~Ahn, P.~S.~Allison, M.~G.~Bagliesi,
  J.~J.~Beatty, G.~Bigongiari, P.~J.~Boyle, T.~J.~Brandt and
  J.~T.~Childers {\it et al.},
  Astropart.\ Phys.\  {\bf 30}, 133 (2008)
  [arXiv:0808.1718 [astro-ph]].

\bibitem{Obermeier:2011wm} A.~Obermeier, M.~Ave, P.~Boyle,
  C.~.Hoppner, J.~Horandel and D.~Muller,
  Astrophys.\ J.\  {\bf 742}, 14 (2011)
  [arXiv:1108.4838 [astro-ph.HE]].

\bibitem{Adriani:2010rc} 
  O.~Adriani {\it et al.}  [PAMELA Collaboration],
  Phys.\ Rev.\ Lett.\  {\bf 105}, 121101 (2010)
  [arXiv:1007.0821 [astro-ph.HE]].

\bibitem{Cholis:2013lwa} 
  I.~Cholis and D.~Hooper,
  Phys.\ Rev.\ D {\bf 89}, no. 4, 043013 (2014)
  [arXiv:1312.2952 [astro-ph.HE]].

\bibitem{Maurin:2010zp} 
  D.~Maurin, A.~Putze and L.~Derome,
  Astron.\ Astrophys.\  {\bf 516}, A67 (2010)
  [arXiv:1001.0553 [astro-ph.HE]].

\end{thebibliography}
\end{document}